\documentclass[preprint,aps,preprintnumbers,nofootinbib,showpacs]{revtex4-1}
\pdfoutput=1
\usepackage{graphicx,color,amsmath,amsfonts,multirow}
\usepackage{epsfig}
\usepackage{rotating}
\usepackage{array}
\usepackage{graphicx}
\usepackage{multirow}
\usepackage{epstopdf}
\usepackage{amssymb}
\usepackage{latexsym}
\def\be{\begin{eqnarray}}
\def\ee{\end{eqnarray}}
\newcommand{\fig}{\begin{figure}}
\newcommand{\ef}{\end{figure}}

\begin{document}
\draft

\topmargin=-1.5cm

\title{\mbox{}\\[10pt]
 Prediction of  Leptonic CP Phase in  $A_4$ symmetric model }
\author{Sin Kyu Kang$^{a}$}
\author{Morimitsu Tanimoto$^{b}$}
\affiliation{$^a$~Institute for Convergence Fundamental Study, School of Liberal Arts, Seoul-Tech, Seoul 139-743, Korea \\ $^b$~Department of Physics, Niigata University, Niigata 950-2181, Japan}


\date{\today}

\begin{abstract}
We consider minimal modifications to tribimaximal (TBM) mixing matrix which accommodate 
non-zero mixing angle $\theta_{13}$ and CP violation.
We derive four possible forms for the minimal modifications to TBM mixing in a model with $A_4$ flavor symmetry by incorporating symmetry breaking terms appropriately. 
We show how  possible values of the  Dirac-type CP phase $\delta_D$ can be predicted with regards to
two neutrino mixing angles in the standard parametrization of the neutrino mixing matrix.
Carrying out numerical analysis based on the recent updated experimental results for neutrino mixing angles, we  predict the values of the CP phase for all possible cases.
We also confront our predictions of the CP phase with the updated fit.

\noindent

\end{abstract}

\maketitle


\section{ Introduction}
Establishing leptonic CP violation (LCPV) is one of the most challenging tasks in future neutrino experiments \cite{cpv-lepton}.
The relatively large value of the reactor mixing angle measured with a high precision in neutrino epxeriments \cite{theta13} has opened up a wide range of possibilities to explore CP violation in the lepton sector.
The LCPV can be induced by the PMNS neutrino mixing matrix \cite{pmns} which contains, in addition to the three angles, a Dirac type CP violating phase in general as it exists in the quark sector, and two extra phases if neutrinos are Majorana particles.
Although we do not yet have compelling evidence for LCPV, the current global fit to available neutrino data indicates nontrivial values of the Dirac-type CP phase  \cite{cp-fit1,cp-fit2}.
In this situation, it must deserve to predict  possible size of LCPV detectable through neutrino oscillations.
From the point of view of {\it calculability}, much attention has been paid to the prediction of the Dirac type LCPV phase with regards to some observables \cite{cp-angle}. 
Recently, it has been shown \cite{kang-kim} that  Dirac-type leptonic CP phase  can be particularly predictable in terms of neutrino mixing angles in the standard parameterization of PMNS mixing matrix \cite{pdg}. 

Before the measurements of the reactor mixing angle, the fit to neutrino data was consistent with
the so-called tribimaximal (TBM) neutrino mixing matrix, $U_0^{\rm TBM}$, which is theoretically well motivated flavor mixing pattern \cite{tribi}.
However, it should be modified to accommodate non-zero reactor mixing angle as well as CP violation.
Although the current neutrino data rule out the exact TBM mixing pattern, it can be regarded
as leading order approximation.
Among various possible modification to $U_{\rm TBM}$, as discussed in \cite{kang-kim}, the minimal modificaton  is useful to predict Dirac type CP phase.
The minimal modification is to multiply $U_0^{\rm TBM}$ by a rotation matrix in the ($i,j$) plane with an angle $\theta$ and a CP phase $\xi$, $U_{ij}(\theta, \xi)$, whose form is given either $U^{\dagger}_{ij}(\theta, \xi) U_0^{\rm TBM}$ or $U_0^{\rm TBM}U_{ij}(\theta,\xi)$ \cite{one-unitary}.
Among them, $U_{23}^{\dagger}(\theta, \xi)U_0^{\rm TBM}$ and
$U_0^{\rm TBM}U_{12}(\theta, \xi)$ are ruled out because 
they lead to zero reactor mixing angle.
So, all possible forms of minimal modification to TBM mixing matrix are as follows:
\begin{eqnarray}
V=\left\{ \begin{array}{l}
     U_{0}^{\rm TBM} U_{23}(\theta, \xi) ~~~\mbox{(Case--A)},\\
     U_{0}^{\rm TBM} U_{13}(\theta, \xi) ~~~\mbox{(Case--B)},\\
     U^{\dagger}_{12}(\theta, \xi) U_{0}^{\rm TBM}  ~~~\mbox{(Case--C)},\\
     U^{\dagger}_{13}(\theta, \xi) U_{0}^{\rm TBM}  ~~~\mbox{(Case--D)}.
     \end{array}\right.
    \label{MTB}
\end{eqnarray}

While the study in \cite{kang-kim} has not accounted for the origin of such modification to $U_0^{\rm TBM}$, in this paper, we first study how such a minimally modified TBM mixing pattern can be achieved in a neutrino model with $A_4$ flavor symmetry by incorporating  $A_4$ symmetry breaking terms appropriately. Then, following \cite{kang-kim}, we investigate how the Dirac type CP phase
can be predicted based on the updated fit results for neutrino mixing angles \cite{cp-fit2}.
As shown later,  comparing with the results obtained in \cite{kang-kim}, the Dirac type CP phase  predicted based on the updated fit results has different implication
particularly at $1 \sigma$ C.L.

\section{Minimal modifications to Tri-bimaximal mixing in $A_4$ symmetric model}
In \cite{ma1}, an $A_4$ symmetric model for neutrino masses and mixing has been proposed to accommodate non-zero mixing angle $\theta_{13}$ on top of TBM mixing. 
Based on the $A_4$ symmetric model, we study how the forms given in Eq.(\ref{MTB}) can be derived 
by incorporating appropriate $A_4$ symmetry breaking terms.
\subsection{Case-A}
As proposed in \cite{ma1}, $A_4$ flavor symmetry allows the charged-lepton mass matrix to be diagonalized by
the Cabibbo-Wolfenstein matrix \cite{CW},
\be
U_{CW}=\frac{1}{\sqrt{3}} \left ( \begin{array}{ccc}
1 & 1 & 1 \\
1 & \omega & \omega^2 \\
1 & \omega^2 & \omega 
\end{array} \right), \label{CW}
\ee
where $\omega=e^{2\pi i/3}$, with three independent eigenvalues, $m_e, m_{\mu}, m_{\tau}$.
This can be realized by the lepton assignments, $L_i=(\nu_i, l_i)\sim \underline{3},~
l^{c}_{1}\sim \underline{1}, ~ l^{c}_{2}\sim \underline{1}^{\prime},  ~l^{c}_{3}\sim \underline{1}^{\prime \prime} $ with 3 Higgs doublets $\Phi_i=(\phi^{0}_i,\phi^{-}_i)\sim  \underline{3}$.
Introducing 6 heavy $A_4$ Higgs singlets and triplet:
\be
\eta_1 \sim \underline{1}, ~\eta_2 \sim \underline{1}^{\prime},~\eta_3 \sim \underline{1}^{\prime \prime},~ \eta_{i(=4,5,6)}\sim \underline{3},
\ee
where $\eta_i=(\eta_i^{++},\eta_i^{+},\eta_i^0)$, 
one can obtain the neutrino mass matrix in the $A_4$ basis \cite{ma1}
\be
M_{\nu}=\left(\begin{array}{ccc}
a+b+c & f & e \\
f & a+\omega b + \omega^2 c & d \\
e & d & a+\omega^2 b + \omega c \end{array} \right ), \label{mass}
\ee
where $a$ comes from $<\eta_1^0>$, $b$ from $<\eta_2^0>$, $c$ from $<\eta_3^0>$,
$d$ from $<\eta_4^0>$, $e$ from $<\eta_5^0>$, $f$ from $<\eta_6^0>$.
To achieve TBM mixing pattern of the neutrino mixing matrix, $A_4$ flavor symmetry should be broken
to $Z_2$ in such a way that $b=c$ and $e=f=0$.
Then, the neutrino mass matrix in the flavor basis where the charged lepton mass matrix is diagonal
is given by
\be
M_{\nu}^{(e,\mu,\tau)}=U^{\dagger}_{CW} M_{\nu}  U^{\ast}_{CW}
=\left(\begin{array}{ccc}
a+(2d/3) & b-(d/3) & b-(d/3) \\
b-(d/3) & b+(2d/3) & a-(d/3) \\
b-(d/3) & a-(d/3)&  b+(2d/3) \end{array} \right ),
\ee
which is diagonalized by the TBM mixing matrix $U_0^{\rm TBM}$.
To achieve non-zero mixing angle $\theta_{13}$ so as to accommodate neutrino data
from reactor experiments, we take $b=c$ and $e=-f \equiv \epsilon\ne 0$ in Eq.(\ref{mass}), and
then the neutrino mass matrix in the $A_4$ basis is given by
\be
M_{\nu}=\left(\begin{array}{ccc}
a+2b & \epsilon & -\epsilon \\
\epsilon &a-b & d \\
-\epsilon & d & a-b \end{array} \right ). \label{mass2}
\ee
In the flavor basis, the neutrino mass matrix can be rewritten as
\be
M_{\nu}^{(e,\mu,\tau)}=\left(\begin{array}{ccc}
 a+(2d/3) & b-(d/3) & b-(d/3) \\
b-(d/3) & b+(2d/3) & a-(d/3) \\
b-(d/3) & a-(d/3)&  b+(2d/3) \end{array} \right )
+\frac{i}{\sqrt{3}}\left(\begin{array}{ccc}
 0& -\epsilon & \epsilon \\
-\epsilon & -2\epsilon & 0 \\
\epsilon & 0 &  2\epsilon \end{array} \right ). \label{mass2}
\ee
Rotating the mass matrix given in Eq.(\ref{mass2}) by TBM mixing matrix, we get
\be
\left(\begin{array}{ccc}
a-b+d & 0 & 0 \\
0 & a+ 2 b & X \\
0 & X & b-a+d 
\end{array}
  \right), \label{massA}
\ee
where $X=\sqrt{2}i \epsilon$ and non-zero entries are complex in general.
It can be easily shown that the mass matrix given by Eq.(\ref{mass2}) can be diagonalized by
\be
V^{\prime}=U_0^{\rm TBM} \left(\begin{array}{ccc}
1 & 0 & 0 \\
0 & \cos\theta & -\sin\theta e^{-i\xi} \\
0 & \sin\theta e^{i\xi} & \cos\theta \end{array} \right )\cdot P_{\beta}, \label{mixing1}
\ee
where $P_{\beta}={\rm Diag}[e^{i\beta_1}, e^{i\beta_2}, e^{i\beta_3}]$.

Now, let us check  testability of the cases in this neutrino model by taking into account the sum-rules among the light neutrino masses \cite{sum-rule}.
In the leading order,  the mass eigenvalues are given by $m^{0}_1=a-b+d,~ m^{0}_2=a+2b, ~m^{0}_3=b-a+d$, and thus we get the mass sum rules
\begin{eqnarray}
 m^{0}_{3}&=&   m^{0}_{2} + m^{0}_{1}, ~~~\mbox{for} ~~a=0,  \nonumber \\
 m^{0}_{1}&=& 2 m^{0}_{2} + m^{0}_{3},~~~\mbox{for}~~ b=0. \label{sumruleA}
\end{eqnarray}
Inclusing the perturbation given by the second matrix in Eq.(\ref{mass2}), we get the following sum rule,
\be
\hat{m}_{2}+\hat{m}_{3}= \hat{m}_2^0+\hat{m}^{0}_3, \label{sum-ruleA2}
\ee
where $\hat{m}_{2}\equiv m_2 e^{-i\xi}, ~\hat{m}_{3}\equiv m_3 e^{i\xi}, ~\hat{m}_{2}^0\equiv m_2^0 e^{-i\xi}, ~\hat{m}_{3}^0\equiv m_3^0 e^{i\xi}$ with $m_{i(=1,2,3)}$ representing the mass eigenvalues obtained by diagonalizing the mass matrix Eq.(\ref{massA}).
Plugging Eq.(\ref{sumruleA}) into Eq.(\ref{sum-ruleA2}), we can get the following sum rules for $\xi=0, ~\pi, ~2\pi$,
\begin{eqnarray}
m_1+ m_2 - m_3 &=& 2~ \delta m_2, ~~~\mbox{for} ~~a=0, \nonumber \\
2 m_2 + m_3 - m_1 &=& \delta m_2, ~~~~~\mbox{for} ~~b=0, \label{sumA}
\end{eqnarray}
where $\delta m_2\equiv m_2 - m^0_2$ and we have used $m_1=m_1^0$.
The sum rules for $\xi=
\pi/2, (3\pi/2)$ are
\begin{eqnarray}
m_1+ m_2 &=& m_3, ~~~~~~\mbox{for} ~~a=0, \nonumber \\
2 m_2 + m_3 - m_1 &=& 3~\delta m_2, ~~~\mbox{for} ~~b=0. \label{sumA2}
\end{eqnarray}

\subsection{Case-B}
To realize the case B, we add the breaking terms $\delta M_{\nu}$ to $M_{\nu}$  in the $A_4$ basis, which is given by
\be
\delta M_{\nu} =\left(\begin{array}{ccc}
g+h & 0 & 0 \\
0 & \omega g + \omega^2 h& 0 \\
0 & 0 & \omega^2 g + \omega h \end{array} \right)
=\left(\begin{array}{ccc}
0& 0& 0 \\
0& A & 0 \\
0& 0 & -A \end{array}
 \right), \label{mass3}
\ee
where $g=-h$ and $A=\sqrt{3}i g$.

Then, the matrix given in Eq.(\ref{mass3}) becomes in the flavor basis as follows:
\be
 \left(\begin{array}{ccc}
0 & g& -g \\
g & \ -g& 0 \\
-g & 0 & g \end{array} \right ). \label{mass4}
\ee
Then, the mass matrix $M_{\nu}+\delta M_{\nu}$ can be diagonalized by
\be
V=U_0^{\rm TBM}  \left(\begin{array}{ccc}
 \cos\theta & 0 &  -\sin\theta e^{-i\xi} \\
0 & 1 & 0 \\
\sin\theta e^{i\xi} & 0 & \cos\theta \end{array} \right )\cdot P_{\beta}.\label{mixing2}
\ee

For the case B, the sum rules at the leading order are the same as Eq.(\ref{sumruleA}).
Including the perturbation given by the second matrix in Eq.(\ref{mass4}), we get the following sum rule
\begin{eqnarray}
\hat{m}_{1}+\hat{m}_{3} = \hat{m}_1^0+\hat{m}^{0}_3, \label{sum-ruleB2}
\end{eqnarray}
where $\hat{m}_{1}\equiv m_1 e^{-i\xi}, ~\hat{m}_{3}\equiv m_3 e^{i\xi}, ~\hat{m}_{1}^0\equiv m_1^0 e^{-i\xi}, ~\hat{m}_{3}^0\equiv m_3^0 e^{i\xi},$ with $m_{i(=1,2,3)}$ representing the mass eigenvalues obtained by diagonalizing the mass matrix $M_{\nu}+\delta M_{\nu}$.
Plugging Eq.(\ref{sumruleA}) into Eq.(\ref{sum-ruleB2}), we can get the following sum rules for $\xi=0,~\pi, ~2\pi$,
\begin{eqnarray}
m_3 - m_2 - m_1=2~\delta m_3, ~~~\mbox{for}~~ a=0, \nonumber \\
m_3+2 m_2 - m_1 = 2~\delta m_3, ~~~\mbox{for}~~ b=0, \label{sumB}
\end{eqnarray}
where $\delta m_3\equiv m_3 - m^0_3$ and we have used $m_2=m_2^0$.
The sum rules for $\xi=\pi/2, (3\pi/2)$ are
\begin{eqnarray}
m_1 - m_3 + m_2=0, ~~~\mbox{for}~~ a=0, \nonumber \\
m_1 - 2 m_2 - m_3 = 0, ~~~\mbox{for}~~ b=0. \label{sumB2}
\end{eqnarray}

\subsection{Case-C}
The case C can be realized by adding the $A_4$ breaking term $\delta M_l$ to the charged lepton mass matrix $M_{l}$ :
\begin{eqnarray}
\delta M_{l}=\left(\begin{array}{ccc}
 g_1 v_1 & g_2 v_1 & 0 \\
g_1 \omega v_2 & g_2 v_2 & 0 \\
g_1 \omega^2 v_3 & g_3 v_3 & 0 
\end{array} \right) . \label{charged1}
\end{eqnarray}
Taking $v_1=v_2=v_3$ and $g_1=g_2=g$, the matrix given by (\ref{charged1}) becomes
\begin{eqnarray}
\delta M_{l}=\left(\begin{array}{ccc}
g v & g v & 0 \\
g \omega v & g v & 0 \\
g \omega^2 v & g v & 0 \end{array} \right)
=U_{CW}\left( \begin{array}{ccc}
0 & g & 0  \\
g & 0 & 0 \\
0 & 0 & 0 \end{array}\right)\sqrt{3} v .
\end{eqnarray}
Due to the addition of $\delta M_{l}$, the PMNS mixing matrix should be changed to
$U^{\dagger}_{12}(\theta,\xi) U_0^{\rm TBM}P_{\beta}$.

For Case C, the sum rules are given by Eq.(\ref{sumruleA}).

\subsection{Case-D}
Similarily, the case D can be achieved by adding the following matrix $\delta M_{l}$ to the charged lepton mass matrix $M_l$:
\begin{eqnarray}
\delta M_{l}=\left(\begin{array}{ccc}
 g_1 v_1 & 0 &g_2 v_1  \\
g_1 \omega^2 v_2 & 0 & g_2 v_2  \\
g_1 \omega v_3 & 0 & g_3 v_3  
\end{array} \right) . \label{charged2}
\end{eqnarray}
Taking $v_1=v_2=v_3$ and $g_1=g_2=g$,  the matrix given in (\ref{charged2}) becomes
\begin{eqnarray}
\delta M_{l}=\left(\begin{array}{ccc}
g v & 0& g v  \\
g \omega^2 v & 0 &  g v \\
g \omega v & 0 &  g v  \end{array} \right)
=U_{CW}\left( \begin{array}{ccc}
0  & 0 & g \\
0 & 0 & 0 \\
g & 0 & 0 \end{array}\right)\sqrt{3} v .
\end{eqnarray}
The addition of $\delta M_{l}$ causes the PMNS mixing matrix changed to
$U^{\dagger}_{13}(\theta,\xi) U_0^{\rm TBM}P_{\beta}$.

For Case D, the sum rules are given by Eq.(\ref{sumruleA}).

\section{Predictions of Dirac-type CP Phase}
Now, let us review how to predict Dirac-type CP phase in PMNS mixing matrix with regards to neutrino mixing angles presented in \cite{kang-kim}.
Multiplying  $V$ given in Eq.(\ref{MTB}) by phase matrices $P_{\alpha}$ and  $P_{\beta}$ that can be arisen from the charged lepton sector and neutrino sector, respectively, we can equate it with  the standard parameterization of the PMN mixing matrix as follows:
\be
&&P_{\alpha}\cdot V \cdot P_{\beta} =
 U^{\rm ST} = U_0^{\rm PMNS}\cdot P_{\phi}\nonumber \\
 &&=
 \left(\begin{array}{ccc}
 c_{12}c_{13} & s_{12}c_{13} & s_{13}e^{-i\delta_D}\\
 -s_{12}c_{23}-c_{12}s_{23}s_{13}e^{i\delta_D} &  c_{12}c_{23}-s_{12}s_{23}s_{13}e^{i\delta_D}&  s_{23}c_{13} \\
 s_{12}s_{23}-c_{12}c_{23}s_{13}e^{i\delta_D} &  -c_{12}s_{23}-s_{12}c_{23}s_{13}e^{i\delta_D}&  c_{23}c_{13} \\
 \end{array}\right)P_{\phi} .
 \label{standard}
 \end{eqnarray}
The equivalence between both parameterizations dictates the following relations,
\begin{eqnarray}
V_{ij}e^{i(\alpha_i + \beta_j)} = U^{\rm ST}_{ij}
= (U_0^{\rm PMNS})_{ij} e^{i\phi_j} ~. \label{rel1}
\end{eqnarray}
\subsection{Case A and  B}
Applying $|V_{13}|=|U_{13}^{\rm ST}|$ and $ |V_{11}/V_{ 12}|=|U_{11}^{\rm ST}/U_{12}^{\rm ST}|$, we obtain the relations 
\be
\sin^2\theta &&= 3 s^2_{13}, \nonumber \\
\cos^2\theta &&= \left\{\begin{array}{c}
                                  2 \tan^2 \theta_{12} ~~({\rm Case A}), \label{rels1} \\
                                  \frac{1}{2} \cot^2\theta_{12}~~({\rm Case B}), \label{rels2} \end{array} \right.
\ee
which lead to the relation between
the solar and  reactor mixing angles, 
\begin{eqnarray}
 s^2_{12}=  \left\{\begin{array}{r}
1-\frac{2}{3(1-s^2_{13})}~~({\rm Case A}),   \\
\frac{1}{3(1-s^2_{13})}~~({\rm Case B}).  \end{array} \right. \label{angle1}
\end{eqnarray}
Those relations indicate that non-zero values of $s^2_{13}$ lead to  $s^2_{12}<1/3$ for Case A and
$s^2_{12} > 1/3$ for Case B .
From $ |V_{23}/V_{ 33}|=|U_{23}^{\rm ST}/U_{33}^{\rm ST}|$, we also get the relations
\be
|\cos\eta| = \left\{\begin{array}{r}
      \frac{c^2_{13}(s_{23}^2-c^2_{23})}{2 s_{13} \sqrt{2-6 s^2_{13}}} ~~({\rm Case A}), \label{phaA} \\
\frac{c_{13}^2 (c^2_{23}-s^2_{23})}{s_{13} \sqrt{2-3s_{13}^2}}~~({\rm Case B}). \label{phaB}
\end{array}\right.
\ee

Now, we demonstrate how to derive $\delta_D$  in terms of neutrino mixing angles
in the standard parametrization. 
From the components of the neutrino mixing matrix  for Case--A, we see that
\begin{eqnarray}
  \frac{V_{23}+V_{33}}{V_{22}+V_{32}}=\frac{V_{13}}{V_{12}}. \label{relA-3}
\end{eqnarray}
From  the relation (\ref{rel1}), we get the relations
  \begin{eqnarray}
  \frac{U^{\rm ST}_{13}}{U^{\rm ST}_{12}} &=&
  \frac{U^{\rm ST}_{23}+U^{\rm ST}_{33} e^{-i(\alpha_3-\alpha_2)}}
          {U^{\rm ST}_{22}+U^{\rm ST}_{32} e^{-i(\alpha_3-\alpha_2)}},
\label{relA1}\\
 \frac{U^{\rm ST}_{3i}}{U^{\rm ST}_{2i}} &=&\frac{V_{3i}}{V_{2i}}e^{i(\alpha_3-\alpha_2)} .
 \label{relA2}
  \end{eqnarray}
Since $V_{21}=V_{31}$, 
\be
e^{i(\alpha_3 - \alpha_2)}=\frac{U^{\rm ST}_{31}}{U^{\rm ST}_{21}}. \label{phase1}
\ee
Plugging Eq.(\ref{phase1}) into Eq.(\ref{relA1}), we finally obtain the relation
  \begin{eqnarray}
  \frac{U^{\rm ST}_{13}}{U^{\rm ST}_{12}}=
 \frac{U^{\rm ST}_{23}U^{\rm ST}_{31}+U^{\rm ST}_{33}U^{\rm ST}_{21}}
         {U^{\rm ST}_{22}U^{\rm ST}_{31}+U^{\rm ST}_{32}U^{\rm ST}_{21}}. \label{AA-final}
  \end{eqnarray}
Notice that the Majorana phases in Eq.(\ref{AA-final}) are cancelled.
Presenting $U^{\rm ST}_{ij}$ explicitly in terms of  the neutrino mixing angles as well as  $\delta_D$, we get the equation for $\delta_D$ as
\begin{eqnarray}
 \cos\delta_D=\frac{-1}{2\tan 2 \theta_{23}}\cdot 
 \frac{1-5 s^2_{13}}{s_{13} \sqrt{2-6 s^2_{13}}}. \label{phaseA}
\end{eqnarray}
 Notice that the imaginary part in Eq. (\ref{AA-final}) is automatically cancelled.

Similarily, we get the relation for Case B,
\be
 \cos\delta_D &=& \frac{1}{2\tan 2 \theta_{23}}\cdot 
 \frac{2-4 s^2_{13}}{s_{13} \sqrt{2-3 s^2_{13}}}. \label{phaseB}
\ee

\subsection{Case C and D}
Applying $|V_{13}|=|U_{13}^{\rm ST}|$ and $ |V_{23}/V_{33}|=|U_{23}^{\rm ST}/U_{33}^{\rm ST}|$, we obtain the relations 
\be
\sin^2\theta &&= 2 s^2_{13}, \nonumber \\
\cos^2\theta &&= \left\{\begin{array}{c}
                                  \tan^2\theta_{23} ~~({\rm Case C}), \label{rels3} \\
                                  \cot^2\theta_{23}~~({\rm Case D}), \label{rels4} \end{array} \right.
\ee
which lead to the relation between
the atmospheric and  reactor mixing angles, 
\begin{eqnarray}
 s^2_{23}=  \left\{\begin{array}{r}
1-\frac{1}{2(1-s^2_{13})}~~({\rm Case C}),   \\
\frac{1}{2(1-s^2_{13})}~~({\rm Case D}). \end{array} \right. \label{angle2}
\end{eqnarray}
Those relations indicate that non-zero values of $s^2_{13}$ lead to  $s^2_{23}<1/2$ for Case C and
$s^2_{23} > 1/2$ for Case D .
From $ |V_{11}/V_{12}|=|U_{11}^{\rm ST}/U_{12}^{\rm ST}|$, we also get the relation
\be
|\cos\eta| = 
      \frac{3c^2_{13}s^2_{12}-1}{2 s_{13} \sqrt{2-4s^2_{13}}}.\label{phaC} 
\ee
We note that  both cases lead to the same relation for $|\cos\eta|$.

Following the same procedures for obtaining Eqs.(\ref{phaseA}, \ref{phaseB}),  we get the relations
\be
 \cos\delta_D = \left\{\begin{array}{r}
\frac{s^2_{13}-(1-3s^2_{12})(1-3 s^2_{13})}{6s_{12}c_{12}s_{13} \sqrt{1-2 s^2_{13}}}~~({\rm Case C}), \label{phaseC}\\
\frac{(1-3s^2_{12})(1-3 s^2_{13})-s^2_{13}}{6s_{12}c_{12}s_{13} \sqrt{1-2 s^2_{13}}}~~({\rm Case D}). \label{phaseD}
\end{array} \right.
\ee
Sustituting experimental values for neutrino mixing angles into Eqs.(\ref{phaseA},\ref{phaseB},\ref{phaseC}), we can estimate
the values of $\delta_D$ for each cases.

\section{Numerical Results}
For our numerical analysis, we take the current experimental data for three neutrino mixing angles as inputs,
which are given at $1\sigma-3\sigma$ C.L., as presented in Ref. \cite{cp-fit2}.
This analysis is, in fact, to update the numerical results for the prediction of $\delta_D$  given in \cite{kang-kim} by taking
the new fit to the data \cite{cp-fit2}.
However, as shown later, the results based on $1 \sigma$ data is completely different from those in \cite{kang-kim}.
Here, we perform numerical analysis and present results only for normal hierarchical neutrino mass spectrum.
It is straight-forward to get numerical results for the inverted hierarchical case, and we anticipate that
the conclusion is not severly changed in the inverted hierarchical case.
Using experimental results for three neutrino mixing angles, we first check if the relations Eqs.(\ref{angle1},\ref{angle2})  hold and then estimate  the values of  $\delta_D$  in terms of neutrino mixing angles for those four cases.

\begin{figure}[h]
\begin{center}
\includegraphics[width=0.49\linewidth]{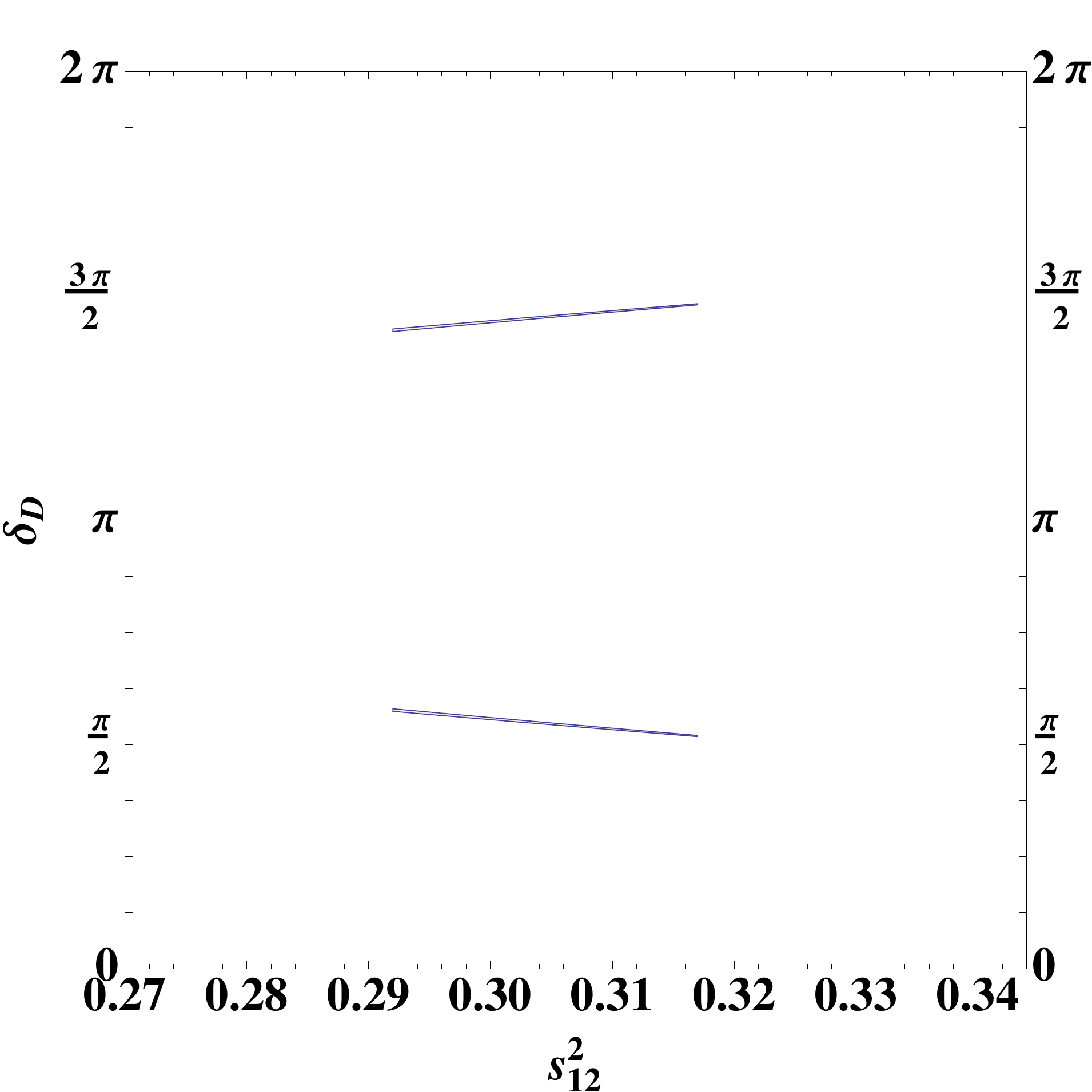}
\end{center}
\caption{Prediction of $\delta_D$ in terms of $s^2_{12}$  for Case C based on $1\sigma$ experimental data.}
\label{fig1}
\end{figure}

\subsection{Results for 1 $\sigma$ C.L.}
Plugging the experimental data for $s^2_{13}$ at 1$\sigma$ C.L. into Eqs.(\ref{angle1},\ref{angle2}), we predict the values of the mixing parameters
$s^2_{12}$ (Case A and B) and $s^2_{23}$ (Case C and D) as follows:
\begin{eqnarray}
s^2_{12}=\left\{\begin{array}{c}
               0.318-0.319 (\rm Case A),\\
               0.340-0.341 (\rm Case B), \end{array} \right. \\
s^2_{23}=\left\{\begin{array}{c}
                 0.488-0.489 (\rm Case C), \\
                0.510-0.511 (\rm Case D). \end{array} \right.
\end{eqnarray}
We see that $s^2_{12}$ and $s^2_{23}$ are very narrowly determined for the $1 \sigma$ region of 
$s^2_{13}$.
Comparing the experimental values of $s^2_{12}$ and $s^2_{23}$  with the above predictions, we see that
only Case C is consistent with experimental results at $1 \sigma$ C.L.

In Fig. \ref{fig1}, we show the prediction of $\delta_D$ in terms of $s^2_{12}$ based on the experimental data at $1\sigma$ C.L.
The upper curve in Fig. \ref{fig1} indicates $1.32 \pi \leq \delta_D \leq 1.52 \pi$ which is consistent with the result of fit for CP phase
$(1.3 \pi \leq \delta_D \leq 1.92\pi)$ shown in \cite{cp-fit2}.
\subsection{Results for 3 $\sigma$ C.L.}
Plugging the experimental data for $\sin^2\theta_{13}$ at 3$\sigma$ C.L. into Eqs.(\ref{angle1},\ref{angle2}), we predict the values of the mixing parameters
$\sin^2\theta_{12}$ (Case A and B) and $\sin^2\theta_{23}$ (Case C and D) as follows:
\begin{eqnarray}
s^2_{12}=\left\{\begin{array}{c}
               0.316-0.321 (\rm Case A),\\
               0.340-0.342 (\rm Case B), \end{array} \right. \\
s^2_{23}=\left\{\begin{array}{c}
                 0.487-0.491 (\rm Case C), \\
                0.509-0.513 (\rm Case D). \end{array} \right.
\end{eqnarray}
Comparing the experimental values of $s^2_{12}$ and $s^2_{23}$  with the above predictions, we see that
they are all consistent with experimental results at $3 \sigma$ C.L.
In particular, the prediction of $s^2_{12}$ for Case B prefers to nearly upper limit of $3 \sigma$ allowed region.
        
\begin{figure}[h]
\begin{center}
\includegraphics[width=0.49\linewidth]{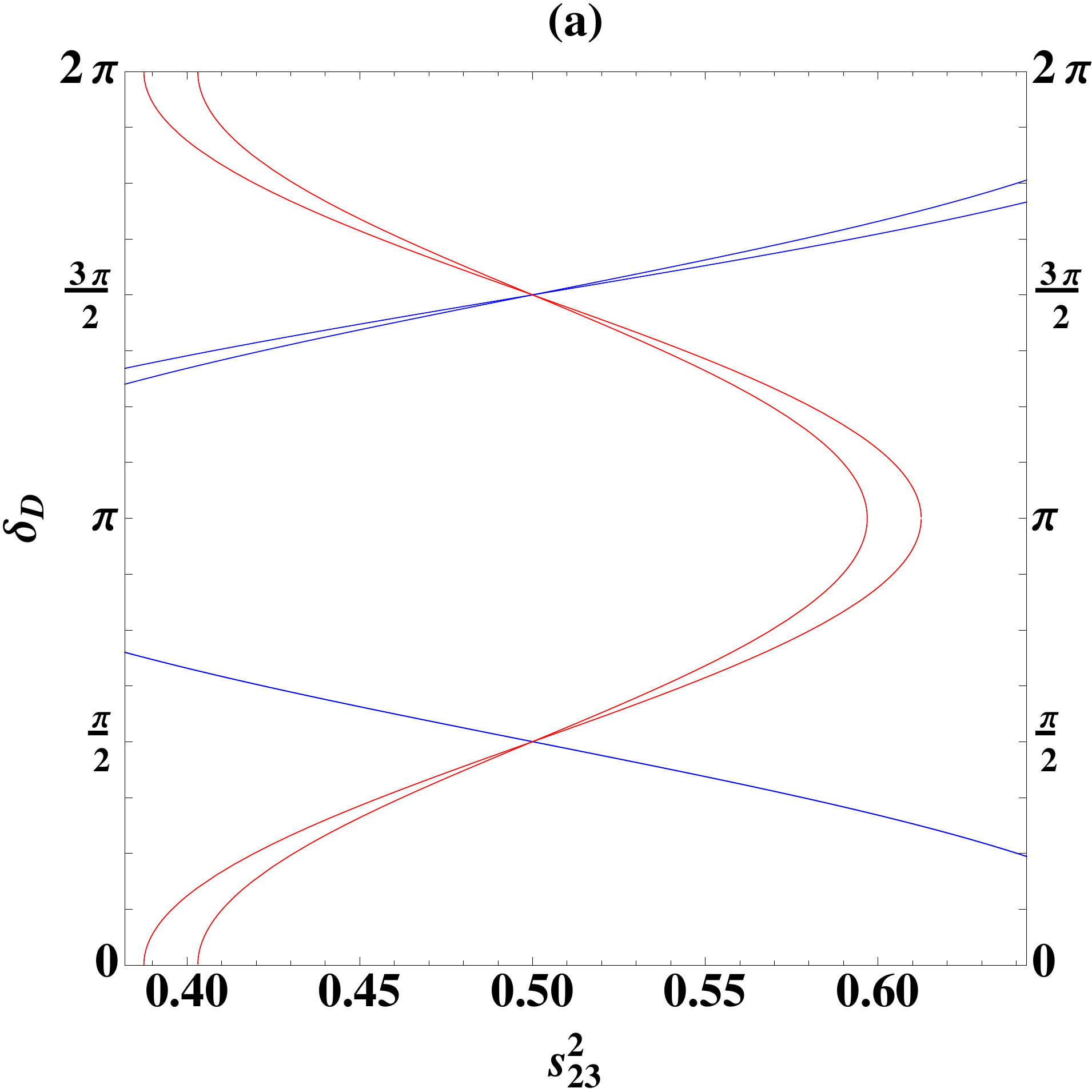}
\includegraphics[width=0.49\linewidth]{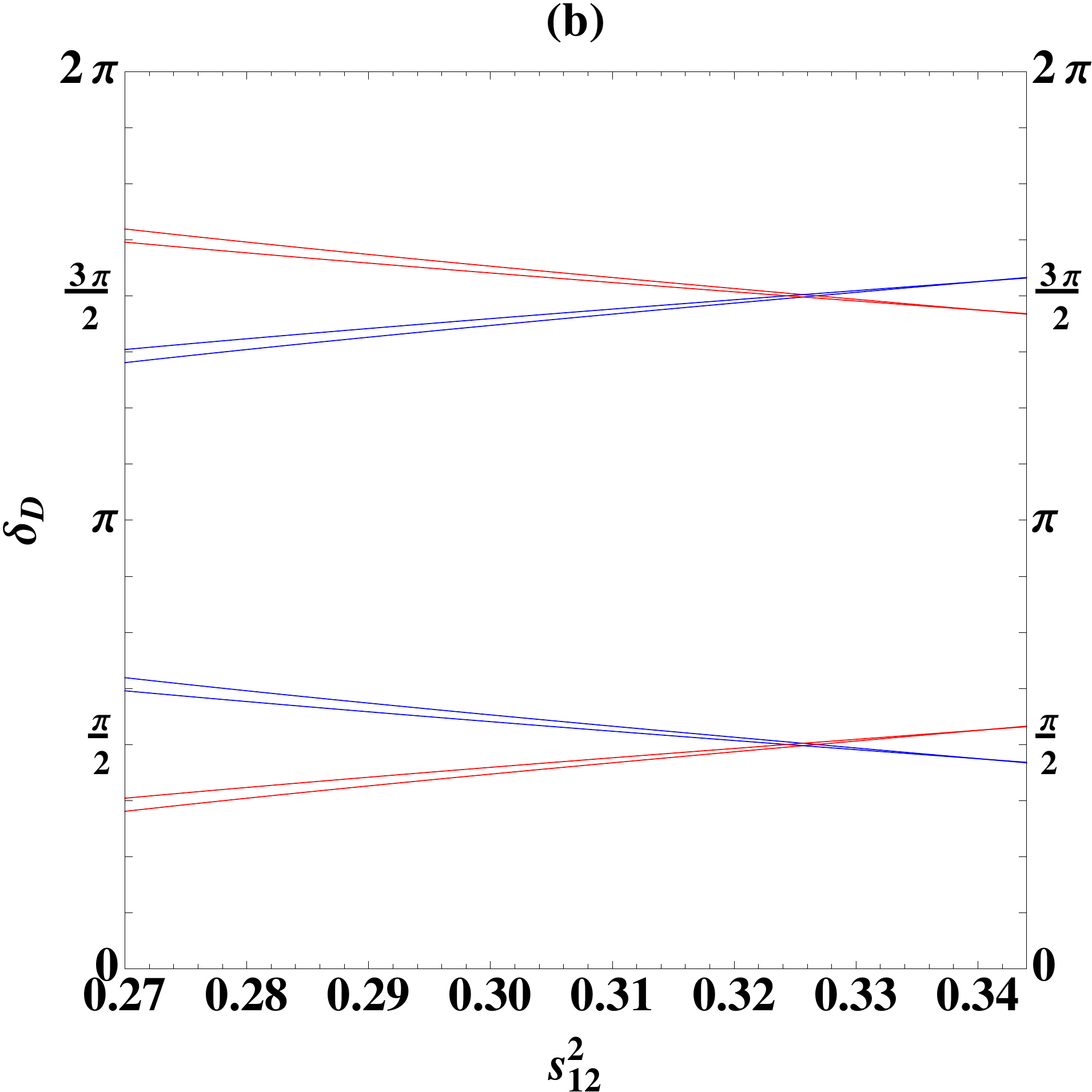}
\end{center}
\caption{Prediction of $\delta_D$ in terms of  (a)  $s^2_{23}$ for Cases A and B, and (b) $s^2_{12}$  for Cases C and D based on $3 \sigma$ experimental data. Regions surrounded by  blue (red) lines correspond to Cases A, C and  (B, D).}
\label{fig2}
\end{figure}

Fig. \ref{fig2} shows the predictions of $\delta_D$ in terms of $s^2_{23}$  ((a): Cases A and B) and $s_{12}^2$ ((b): Cases C and D)   based on the corresponding experimental data given at $3\sigma$ C.L.
Regions surrounded by  blue and red lines correspond to Cases (A, C) and  (B, D), respectively.
The width of each bands implies the variation of the other mixing angles, $s_{12}^2$ (Cases A and B) and $s_{23}^2$ (Cases C and D).  We see that  almost maximal $\delta_D\sim \pi/2, 3\pi/2$ can be achieved by $s^2_{23}\sim 0.5$  for Cases--A, B and by $s^2_{12}\sim0.325$ for Cases--C,D .
The values around $3\pi/2$ is consistent with the current fit of the Dirac type CP phase \cite{cp-fit2}.

Comparing those results with the corresponding ones presented in \cite{kang-kim}, we see that
the shapes of the curves in each cases are nearly unchanged, but the widths of each bands get much narrower. The allowed regions of $s^2_{12}$ above 0.344  for Cases C and D are excluded in the updated analysis.

\section{Conclusion}
As a summary, we have considered how non-zero mixing angle $\theta_{13}$ and CP violation
can be acommodated in a model with $A_4$ flavor symmetry by incorporating symmetry breaking terms appropriately. The four possible forms of neutrino mixing matrix we considered are minimal modifications to TBM mixing matrix and factorized by TBM mixing form and an unitary mixing matrix with an angle and a CP phase corresponding to a rotation in a plane.
We have shown that  possible size of the  Dirac-type CP phase $\delta_D$ can be predicted with regards to
two neutrino mixing angles in the standard parametrization of the neutrino mixing matrix.
This has been achieved by equating  one of minimally modified TBM mixing matrix with the standard parametrization of the PMNS one.
Based on the current fit results for the neutrino mixing angles and CP phase, we have seen that
the neutrino mixing matrix corresponding to Case C is consistent with the current fit data
at $1\sigma $ C.L. whereas others are not so. 
This result is different from that in \cite{kang-kim}.
Extending the anlyais to $3 \sigma$ C.L., all cases are
consistent with the current fit data.
We have presented the numerical results for the predictions of $\delta_D$ in terms of either $s^2_{12}$ or $s^2_{23}$
for those cases.

%
%
%
%

\noindent{\bf Acknowledgments}

The work of S.K.K. was supported by the NRF
grant funded by Korea government of the MEST (No. 2011-0029758).


\end{document}